\documentclass[preprint,showpacs,preprintnumbers,amsmath,amssymb,graphicx]{revtex4}
\usepackage{amsmath,amsfonts,amssymb}
\usepackage{epsfig}
\def\bit{\begin{itemize}}
\def\eit{\end{itemize}}
\def\ben{\begin{enumerate}}
\def\een{\end{enumerate}}
\def\bed{\begin{description}}
\def\eed{\end{description}}

\def\k{\kappa}
\def\l{\lambda}

\def\q{\quad}

\def\half{\frac{1}{2}\,}
\def\third{\frac{1}{3}\,}
\def\quart{\frac{1}{4}\,}
\def\lsim{\raise0.3ex\hbox{$<$\kern-0.75em\raise-1.1ex\hbox{$\sim$}}}
\def\gsim{\raise0.3ex\hbox{$>$\kern-0.75em\raise-1.1ex\hbox{$\sim$}}}

\let\jnfont=\rm
\def\NPB#1,{{\jnfont Nucl.\ Phys.\ B}{\bf #1},}
\def\PLB#1,{{\jnfont Phys.\ Lett.\ B}{\bf #1},}
\def\EPJC#1,{{\jnfont Eur.\ Phys.\ Jour.\ C}{\bf #1},}
\def\PRD#1,{{\jnfont Phys.\ Rev.\ D}{\bf #1},}
\def\PRL#1,{{\jnfont Phys.\ Rev.\ Lett.\ }{\bf #1},}
\def\MPLA#1,{{\jnfont Mod.\ Phys.\ Lett.\ A}{\bf #1},}
\def\JPG#1,{{\jnfont J.\ Phys.\ G}{\bf #1},}
\def\CTP#1,{{\jnfont Commun.\ Theor.\ Phys.\ }{\bf #1},}
\def\JHEP#1,{{\jnfont JHEP }{\bf #1},}
\def\NPPS#1,{{\jnfont Nucl.\ Phys.\ Proc.\ Suppl.\ }{\bf #1},}

\def\beq{\begin{equation}}
\def\eeq{\end{equation}}
\def\bea{\begin{eqnarray}}
\def\eea{\end{eqnarray}}
\newcommand{\ba}{\begin{array}}
\newcommand{\ea}{\end{array}}
\def\nn{\nonumber}

\begin{document}

\title{Higgs boson mass in NMSSM with right-handed neutrino}

\author{Wenyu Wang$^1$, Jin Min Yang$^2$, Lin Lin You$^1$}

\affiliation{
$^1$ Institute of Theoretical Physics, College of Applied Science,
     Beijing University of Technology, Beijing 100124, China\\
$^2$ State Key Laboratory of Theoretical Physics,
     Institute of Theoretical Physics, Academia Sinica,
              Beijing 100190, China ~ \vspace{0.5cm} }

\begin{abstract}
In order to have massive neutrinos, the right-handed
neutrino/sneutrino superfield ($N$) need to be introduced in
supersymmetry. In the framework of NMSSM (the MSSM with a singlet
$S$) such an extension will dynamically lead to a TeV-scale Majorana
mass for the right-handed neutrino through the $SNN$ coupling when
$S$ develops a vev (the free Majorana mass term is forbidden by
 the assumed $Z_3$ symmetry).
Also, through the couplings $SNN$ and $SH_uH_d$, the SM-like Higgs boson
(a mixture of $H_u$, $H_d$ and $S$) can naturally couple with the right-handed
neutrino/sneutrino. As a result, the TeV-scale right-handed
neutrino/sneutrino may significantly contribute to the Higgs boson mass.
Through an explicit calculation, we find that the Higgs boson mass
can indeed be sizably altered by the right-handed neutrino/sneutrino.
Such new contribution can help to push up the SM-like Higgs boson mass
and thus make the NMSSM more natural.
\end{abstract}
\maketitle

\section{introduction}
Supersymmetry (SUSY) \cite{susy,susyhierarchy} gives a natural
solution to the hierarchy problem suffered by the Standard Model
(SM). Also, it provides a good dark matter candidate and realizes
the gauge coupling unification. Among the SUSY models, the Minimal
Supersymmetric Standard Model (MSSM) \cite{mssm} has been
intensively studied. However, the recently discovered Higgs-like
boson around 125 GeV caused a problem for this model, i.e., a 125
GeV Higgs boson requires a heavy stop or a large tri-linear coupling
$A_t$ and thus incurs the little hierarchy problem. Besides, the
MSSM suffers from the $\mu$-problem \cite{muew}.

It is remarkable that both the little hierarchy problem and the
 $\mu$-problem can be solved in the Next-to-Minimal Supersymmetric
Standard Model (NMSSM) \cite{NMSSM} in which an additional gauge
singlet $S$ is introduced (in fact, the NMSSM was proposed even earlier
than the MSSM \cite{fayet}). In this model the $\mu$-problem is solved
by the dynamical generation of the $\mu$-term through
the coupling $SH_uH_d$ when $S$ develops a vev, while
the little hierarchy problem is solved by the generation
of an extra tree-level mass term for the SM-like Higgs boson
(thus the stop mass or $A_t$ is no longer required to be unnaturally
large).

Note that in order to have massive neutrinos, right-handed
neutrino/sneutrino superfield(s) ($N$) need to be introduced in SUSY
models. For the NMSSM with such right-handed neutrino/sneutrino
field(s) \cite{Cerdeno:2008ep}, some intriguing features are
present. Due to  the assumed $Z_3$ symmetry, the free Majorana mass
term for the right-handed neutrino is forbidden in the
superpotential. Instead, a TeV-scale Majorana mass for the
right-handed neutrino is dynamically generated through the $SNN$
coupling when $S$ develops a vev (note that such a TeV-scale
majorana mass is too low for the see-saw mechanism and thus the
neutrino Yukawa couplings $H_u L N$ must be very small). In the same
way, a TeV-scale mass for the right-handed sneutrino can also be
generated, which can serve as a good dark matter candidate
\cite{Cerdeno:2009dv}. Further, through the couplings $SNN$ and
$SH_uH_d$, the SM-like Higgs boson (a mixture of $H_u$, $H_d$ and
$S$) can naturally couple with the right-handed neutrino/sneutrino.
As a result, the TeV-scale right-handed neutrino/sneutrino may
significantly contribute to the Higgs boson mass (in the MSSM with
split SUSY, the right-handed neutrino/sneutrino can also make
sizable contribution to the Higgs mass, as studied in
\cite{cao-split}). In this paper we will perform an explicit
calculation for such contribution.

This work is organized as follows.
In Sec. \ref{sec:model_method} we present the spectrum and
couplings for the Higgs boson and right-handed neutrino/sneutrino.
In Sec. \ref{sec:renm} the renormalization scheme is described.
Numerical results and discussions are given in Sec. \ref{sec:num}.
Finally, we give a summary in Sec.\ref{sec:sum}.

%%%%%%%%%%%%%%%%%%%%%%%%%%%%%%%%%%%%%%%%%%%%%%%%%%%%%%%%%%%%%%%%%%%
\section{Higgs and right-handed neutrino/sneutrino in NMSSM}
\label{sec:model_method}
\subsection{Model description}
The NMSSM with a right-handed neutrino superfield $N$ has a superpotential
given by
\begin{eqnarray}
&&  W = W_{\rm NMSSM} + \lambda_N S N N + y_N H_u \cdot L N, \nn\\
&&  W_{\rm NMSSM} = Y_u H_u \cdot Q u_R - Y_d H_d \cdot Q d_R
  - Y_e H_d \cdot L e_R
  + \lambda S H_u \cdot H_d + \frac{1}{3}\kappa S^3 ,
  \label{superpotential}
\end{eqnarray}
where the flavor indices are omitted and the dot denotes the
$SU(2)_L$ antisymmetric product. Since a global $Z_3$ symmetry is
imposed, there are no supersymmetric mass terms (like $H_uH_d$, $NN$
or $SS$) in the superpotential. Note that in this model we impose R-parity
and thus the terms $NNN$ and $SSN$ are forbidden.
As a result, the sneutrino-Higgs mixing is avoided and also
there is no vev for the right-handed sneutrino (we will show how to get the
globle minimum in the following).
Although a bare Majorana mass term $NN$ is forbidden in the superpotential,
a TeV-scale Majorana mass can be generated through the coupling $SNN$
when $S$ develops a non-zero vev ($v_s$). Such a TeV-scale Majorana
mass is too small for the conventional see-saw mechanism and thus
the Yukawa coupling $y_N H_u L N$ should be very small ($y_N\ll 1$).
Note that here we introduce only one right-handed neutrino superfield
to illustrate its effects on the Higgs mass.
In order to explain the neutrino masses and mixings,
more right-handed neutrino superfields need to be introduced,
each of which will contribute to the Higgs mass.
In this case, the calculation method is same as in our calculation,
but the total effects may be more sizable due to more free parameters.

The soft SUSY breaking terms for Higgs and right-handed sneutrino
are given by (hereafter we use $N$ and $\tilde N$ to denote
respectively right-handed neutrino and sneutrino)
\begin{eqnarray}
  -{\cal L}_{\rm soft}
 &=& M_{H_u}^{2}|H_{u}|^{2}+M_{H_d}^{2}|H_{d}|^{2}+M_{s}^{2}|S|^{2}
 + (\lambda A_{\lambda}H_{u}\cdot H_{d}S+\frac{\kappa}{3}A_{\kappa}S^{3}+h.c.)\nn\\
 && +M_{\tilde N}^2 |\tilde N|^2 +(\lambda_N A_N S \tilde N \tilde N + h.c.)
  \label{softterm}
\end{eqnarray}
Here we neglected the mixing
between left-handed and right-handed sneutrinos
because the mixing is assumed to be suppressed by $y_N$.
In the following we briefly discuss the neutral Higgs neutrino sectors.

\subsection{The neutral Higgs sector}
From Eq. (\ref{superpotential}) and Eq. (\ref{softterm})
we get the Higgs potential
\bea
V & = & \l^2 (|H_u|^2|S|^2 + |H_d|^2|S|^2 + |H_u \cdot  H_d|^2) +
\k^2|S^2|^2 \nn \\
&& + \l\k (H_u \cdot H_dS^*S^* + \mathrm{h.c.})
+ \quart g^2 (|H_u|^2 - |H_d|^2)^2 \nn \\
& & + \half g_2^2 |H_u^+
(H_d^0)^* + H_u^0 (H_d^-)^*|^2
 + M_{H_u}^2|H_u|^2 + M_{H_d}^2|H_d|^2 + M_S^2|S|^2 \nn \\
& &+ (\l A_\l H_u \cdot
H_d S + \third \k A_\k\ S^3 + \mathrm{h.c.})
\label{vform}
\eea
where $g^2 = (g_1^2 + g_2^2)/2$ with $g_1$ and $g_2$ being the SM gauge
coupling constants.
Assuming $H_u$, $H_d$ and $S$ get vevs such that
\beq
H_u^0 = v_u + \frac{{\rm Re}(H_R^0) + i {\rm Im}(H_u^0)}{\sqrt{2}} , \q
H_d^0 = v_d + \frac{{\rm Re}(H_d^0) + i {\rm Im}(H_d^0)}{\sqrt{2}} , \q
S = v_s + \frac{S_R + i S_I}{\sqrt{2}}
\eeq
we can get the mass terms for the Higgs fields, which are presented
in \cite{NMSSMTools}.
Here we only show the conventions and give some brief comments:
\begin{enumerate}

\item The mass matrix for the CP-even neutral Higgs
is obtained from the real components of the Higgs fields.
In the basis $h^{bare} = [{\rm Re}(H_u^0), {\rm Re}(H_d^0), S_R]$
and using the minimization equations to eliminate the soft masses,
one obtains three CP-even states (ordered in mass)
\beq
h_i = S_{ij} h^{bare}_j
\label{Sij}
\eeq
with an orthogonal rotation $S_{ij}$.

\item The mass matrix for the CP-odd neutral Higgs
is obtained form the imaginary components of the
Higgs fields $[{\rm Im}(H_u^0), {\rm Im}(H_d^0), S_I]$.
Its diagonalization is performed in two steps.
First, one rotates it into a basis ($A, S_I, G$)
where $G=-\sin\beta {\rm Im}(H_u^0)+\cos\beta {\rm Im}(H_d^0)$
is a massless Goldstone mode ($\tan\beta = v_u/v_d$ is the ratio of
the vevs of the two Higgs doublets). Dropping the Goldstone mode, the
remaining $2 \times 2$ mass matrix $M^2_p$ in the basis ($A, S_I$)
can be diagonalized by an orthogonal $2 \times 2$ matrix $P_{ij}$
into two physical CP-odd states $a_i$ (ordered in mass):
\bea
a_1 &=& P_{11} A + P_{12} S_I, \nn \\
a_2 &=& P_{21} A + P_{22} S_I .
\label{Pij}
\eea

\item The neutralino mass matrix ${\cal M}_N$ in the basis
$\psi^0 = (-i\l_1 , -i\l_2, \psi_u^0, \psi_d^0,\psi_s)$ can be
diagonalized by an rotation matrix $N_{ij}$. Then one obtains five
eigenstates (ordered in mass) $\chi^0_i = N_{ij} \psi^0_j$.
\end{enumerate}

\subsection{Right-handed neutrino/sneutrino sector}
Since there is no Dirac mass term here, the mass spectrum
of the right-handed neutrino sector is very simple.
Denoting $\tilde N=R+iM$, there are only one
CP-even right-handed sneutrino (denoted as $R$)
and one CP-odd right-handed sneutrino (denoted as $M$).
The right-handed neutrino is denoted as $N$.
From Eq. (\ref{superpotential}) and Eq. (\ref{softterm})
we can get the spectrum as
\bea
m_R^2 &=& 4\l_N^2 v_s^2 + M_{\tilde N}^2 + 2\l_N v_s A_N
+ 2\l_N(\k v_s^2 - \l v_u v_d)\nn\\
m_M^2 &=& 4\l_N^2 v_s^2 + M_{\tilde N}^2 - 2\l_N v_s A_N
- 2\l_N(\k v_s^2 - \l v_u v_d) \nn\\
m_N &=& 2\l_N v_s. \label{rnspc}
\eea
With the above spectrum we can get the couplings between
the Higgs and the right-handed neutrino/sneutrino.
Note that in our numerical study we require $M_R^2$ and $M_M^2$
be positive, and, as a result, the global minimum of the scalar
potential locates at the zero point of the right-handed
sneutrino field (the right-handed sneutrino has no vev and
thus R-parity is preserved).
In the following we list the couplings which
will be used in our later calculations:
\begin{eqnarray}
&& V_{h_iRR}=\sqrt{2}\lambda_{N}\lambda\left(v_{u}S_{j2}+v_{d}S_{j1}\right)
- \sqrt{2}\left(2\lambda_{N}\kappa v_{s}
+4\lambda_{N}^{2}v_{s}+\lambda_{N}A_{N}\right)S_{j3},\\
&& V_{h_iMM}=- \sqrt{2}\lambda_{N}\lambda\left(v_{u}S_{j2}+v_{d}S_{j1}\right)
+ \sqrt{2} \left(2\lambda_{N}\kappa v_{s}
-4\lambda_{N}^{2}v_{s}+\lambda_{N}A_{N}\right)S_{j3},\\
&&V_{h_ih_j RR}=-\lambda_{N}\left[2\kappa S_{j3}S_{i3}
-\lambda(S_{j1}S_{i2}+S_{i1}S_{j2})\right]
-4\lambda_{N}^{2}S_{j3}S_{i3},\\
&&V_{h_ih_j MM}=\lambda_{N}\left[2\kappa S_{j3}S_{i3}
-\lambda(S_{j1}S_{i2}+S_{i1}S_{j2})\right]
-4\lambda_{N}^{2}S_{j3}S_{i3},\\
&&V_{a_i RM} = -2\l_N(-\l v \cos2\beta P_{i1}/\sqrt2 +\sqrt2\k v_s P_{i1})
+ \sqrt2\l_N A_N P_{i2},\\
&&V_{a_ia_j RR} = 2\l_N(\l \sin\beta\cos\beta P_{i1}P_{j1}+\k P_{i2}P_{j2})-4\l_N^2P_{i2}P_{j2},\\
&&V_{a_ia_j MM} = -2\l_N(\l \sin\beta\cos\beta P_{i1}P_{j1}+\k P_{i2}P_{j2})-4\l_N^2P_{i2}P_{j2},\\
&& V_{h_i NN}=-\sqrt{2}\l_NS_{i3}~~~~~~~~
V_{a_i NN}=\sqrt{2}i\l_NP_{i2}\gamma^{5},\\
&&V_{\chi_i R N}=-\l_N\frac{N_{i5}}{2\sqrt{2}}~~~~~~~~~~
V_{\chi_i M N}=\l_N\frac{iN_{i5}\gamma^{5}}{2\sqrt{2}}.
\end{eqnarray}

%%%%%%%%%%%%%%%%%%%%%%%%%%%%%%%%%%%%%%%%%%%%%%%%%%%%%%%%%%%%%%%%
\section{renormalization scheme}
\label{sec:renm} To calculate the neutrino/sneutrino contribution to
the Higgs mass, we must calculate the one-loop Higgs propagator and
choose a renormalization scheme. Here we follow \cite{Ender:2011qh}
and choose the mixed renormalization scheme (other schemes give
similar results). We choose the following parameter set \beq
M_Z,~M_W,~M_{H^\pm},~e,
\underbrace{t_{H_u},~t_{H_d},~t_{H_s},}_{\mbox{on-shell
 scheme}}
\underbrace{\tan \beta, ~\lambda, ~v_s, ~\kappa,
~A_\kappa}_{\overline{\mbox{DR}} \mbox{ scheme}} \;,\label{rescheme}
 \eeq where
$t_{H_u},t_{H_d},t_{H_s}$ are the tadpoles of the CP-even Higgs
fields. Since we concentrate on the right-handed nuetrino/sneutrino
contributions, the input parameters from the gauge interaction part
need not be renormalized. For the parameters which need
renormalization, we replace them by the renormalized ones plus the
corresponding counterterms: \bea
\begin{array}{ll}
 t_{H_u} \to t_{H_u} + \delta t_{H_u}, & ~~\tan \beta \to  \tan \beta +\delta \tan \beta\\
 t_{H_d} \to t_{H_d} + \delta t_{H_d}, &  ~~ \lambda \to \lambda + \delta \lambda\\
 t_{H_s} \to t_{H_s} + \delta t_{H_s}, &  ~~ \k \to \k+\delta\k\\
 v_s \to v_s + \delta v_s,          &  ~~  A_\kappa \to  A_\kappa +\delta A_\kappa~.
\end{array}\label{conterm}
\eea In the following we will show how to determine the counter
terms in the mixed renormalization scheme.

%%%%fig.1 %%%%%%%%%%%%%%%%%%%%%%%%%%%%%%%%%%%%%%%%%%%
\begin{figure}[hbtp]
\epsfig{file=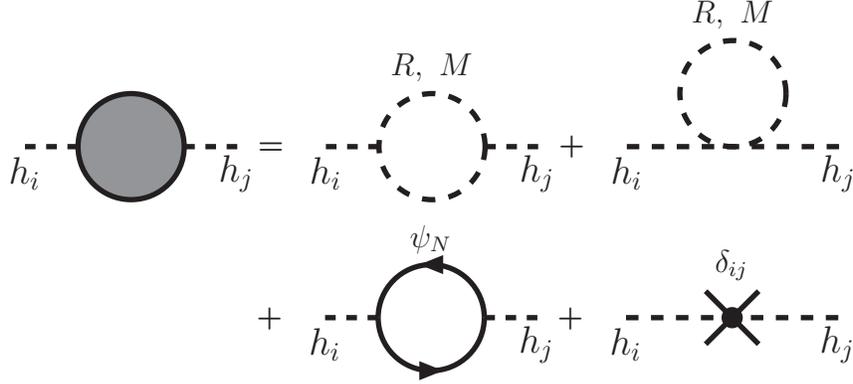}
\caption{Feynman diagrams for the two-point renormalized
Higgs functions.}\label{fig1}
\end{figure}
%%%%%%%%%%%%%%%%%%%%%%%%%%%%%%%%%%%%%%%%%%%%%%%%%%%%%%%
First, the Higgs doublet and singlet fields are replaced by the
renormalized ones:
\begin{alignat}{2}\nonumber
H_u &\to \sqrt{Z_{H_u}} \,H_u  \,&=& \, \left(1 + \frac{1}{2} \delta
  Z_{H_u}\right)  H_u\\
H_d &\to \sqrt{Z_{H_d}} \, H_d  \,&=& \, \left(1 + \frac{1}{2} \delta
  Z_{H_d} \right)  H_d \label{eq:fieldren}\\\nonumber
S &\to \sqrt{Z_{S}} \, S  \,&=& \, \left(1 + \frac{1}{2} \delta Z_{S} \right)  S \;.
\end{alignat}
Then the renormalized two-point functions can be obtained
from the Feynman diagrams shown in Fig. \ref{fig1}
\bea
\hat{\Sigma}_{H_iH_j}(k^2) &=& S_{ik}\; S_{jl}\;\hat{\Sigma}_{kl}^S(k^2)
\qquad  (i,j,k,l = 1,2,3),\\
\hat{\Sigma}_{A_iA_j}(k^2) &=& P_{ik}\;P_{jl}\;
\hat{\Sigma}_{kl}^P(k^2) \qquad (i,j,k,l = 1,2), \eea where $S_{ij}$
and $P_{ij}$ are the matrix elements defined in Eqs.(\ref{Sij}) and
(\ref{Pij}). The renormalization condition can be set as \beq \delta
Z_{H_i H_i} = -\left. \frac{\partial \Sigma_{H_i H_i}
(k^2)}{\partial k^2} \right|_{k^2=(M^{(0)}_{H_i}
)^2}^{\scriptsize{\mbox{div}}} \; \qquad (i=1,2,3) \;, \eeq where
$M_{H_i}^{(0)}$ denotes the corresponding tree-level Higgs mass, and
'div' shows that we chose the $\overline{\text{DR}}$ renormalization
scheme which means that in the field renormalization only the
divergent part $\Delta = 2/(4 - D) - \gamma_E + \ln(4 \pi)$
($\gamma_E$ is the Euler constant)  is kept. The field
renormalization constants $\delta Z_{H_d}, \delta Z_{H_d}, \delta
Z_S$ are obtained by solving the equations \beq \delta Z_{H_i H_i} =
|S_{i1}|^2 \delta Z_{H_d} + |S_{i2}|^2 \delta Z_{H_u} + |S_{i3}|^2
\delta Z_{S} \; \qquad (i =1,2,3) \; . \label{eq:renconst} \eeq We
use the field renormalization constants to determine the conterterms
listed in Eq. (\ref{conterm}). The detailed calculations are
lengthy. In the following we only present the final results and give
some necessary comments.
\begin{enumerate}
\item Tadpole parameters:\\
The tadpole parameters are determined by the condition that
they vanish after the renormalization. The Feynman diagrams
are shown in Fig. \ref{fig2} and the counter terms are determined by
\beq
\delta t_{H_i} = S_{ji} \; t^{(1)}_{h_j}~~~~
(i=u,d,s, ~~j=1,2,3) \;.
\eeq
where $t^{(1)}_{h_j}$ denote the one-loop Higgs tadpoles.
%%%%fig.2 %%%%%%%%%%%%%%%%%%%%%%%%%%%%%%%%%%%%%%%%%%%
\begin{figure}[hbtp]
\epsfig{file=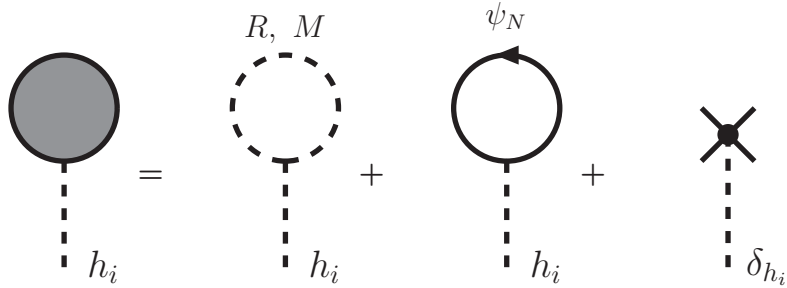}
\caption{Feynman diagram for the Higgs tadpoles.}\label{fig2}
\end{figure}
%%%%%%%%%%%%%%%%%%%%%%%%%%%%%%%%%%%%%%%%%%%
\item The parameter $\tan\beta$:\\
\beq
\delta \tan \beta =
\left[ \frac{\tan \beta }{2} (\delta Z_{H_u} - \delta Z_{H_d})
\right]_{\mbox{\scriptsize{div}}} \;.
\eeq
\item The coupling $\l$:\\
 \begin{align}
 \nonumber
 \delta \lambda = \frac{e^2}{4 \lambda M_W^2 s_W^2}\Bigl[&\Sigma_{P,11}
  (M^2_{P,11}) \Bigr]_{\scriptsize{\mbox{div}}} \label{eq:dellambda} \quad .
\end{align}
The self-energy $\Sigma_{P,11}$ is obtained from the
self-energies in the mass eigenstate basis $\Sigma_{A_iA_j}$
($i,j=1,2,3$) through
\beq
\Sigma_{P,11} = P_{i1} \, \Sigma_{A_i A_j} \, P_{j1} \quad \;.
\eeq
\item The singlet Higgs vev $v_s$:\\
\beq
\delta v_s = \left.- v_s \frac{\delta \lambda}{\lambda}\right|_{\mbox{\scriptsize{div}}} \;,
\label{eq:delvs}
\eeq
\item  The coupling $\kappa$:\\
$\k$ is renormalized through the neutralino renormalization whose
diagrams are shown in Fig. \ref{fig3}. Note that we have different
conventions of vev and thus the formula is a little different from
Ref. \cite{Ender:2011qh}. \beq \delta \kappa = \frac{1}{2 v_s}
\delta ({\cal M}_N)_{55} - \kappa \frac{\delta v_s}{v_s} \;.
\label{eq:delkap} \eeq
%%%%fig.3 %%%%%%%%%%%%%%%%%%%%%%%%%%%%%%%%%%%%%%%%%%%
\begin{figure}[hbtp]
\hspace{50mm}\epsfig{file=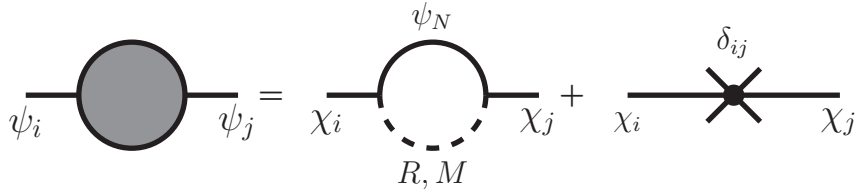}\\ \ \\
\caption{Feynman diagrams for the renormalized two-point neutralino
functions.} \label{fig3}
\end{figure}
%%%%%%%%%%%%%%%%%%%%%%%%%%%%%%%%%%%%%%%%%%%
\item Tri-linear coupling $A_\kappa$:\\
$A_\kappa$ is renormalized by the CP-odd Higgs element $M^2_{P,22}$
and is given by
\bea
\delta A_\kappa = \Bigl[- \frac{1}{3 \kappa v_s}
\bigl[\Sigma_{P,22} \Bigl(M^2_{P,22}\Bigr) - \delta f\bigr]
- A_\kappa \bigl[\frac{\delta \kappa}{\kappa} + \frac{\delta
  v_s}{v_s}\bigr]  \Bigr]_{\scriptsize{\mbox{div}}} \;,
\eea
where the fuction $f$ can be found in Ref. \cite{Ender:2011qh}.
\end{enumerate}
After the determination of the counterterms, we put these terms into
the Higgs mass matrix which is shown in the Appendix. Also, by
adding the loop contribution to the Higgs mass matrix, we can get
the mass correction for the Higgs boson.

%%%%%%%%%
\section{numerical results}\label{sec:num}
\subsection{The right-handed neutrino/sneutrino correction
to the Higgs boson mass}
%%%%fig.4 %%%%%%%%%%%%%%%%%%%%%%%%%%%%%%%%%%%%%%%%%%%
\begin{figure}[hbtp]
\scalebox{0.48}{\epsfig{file=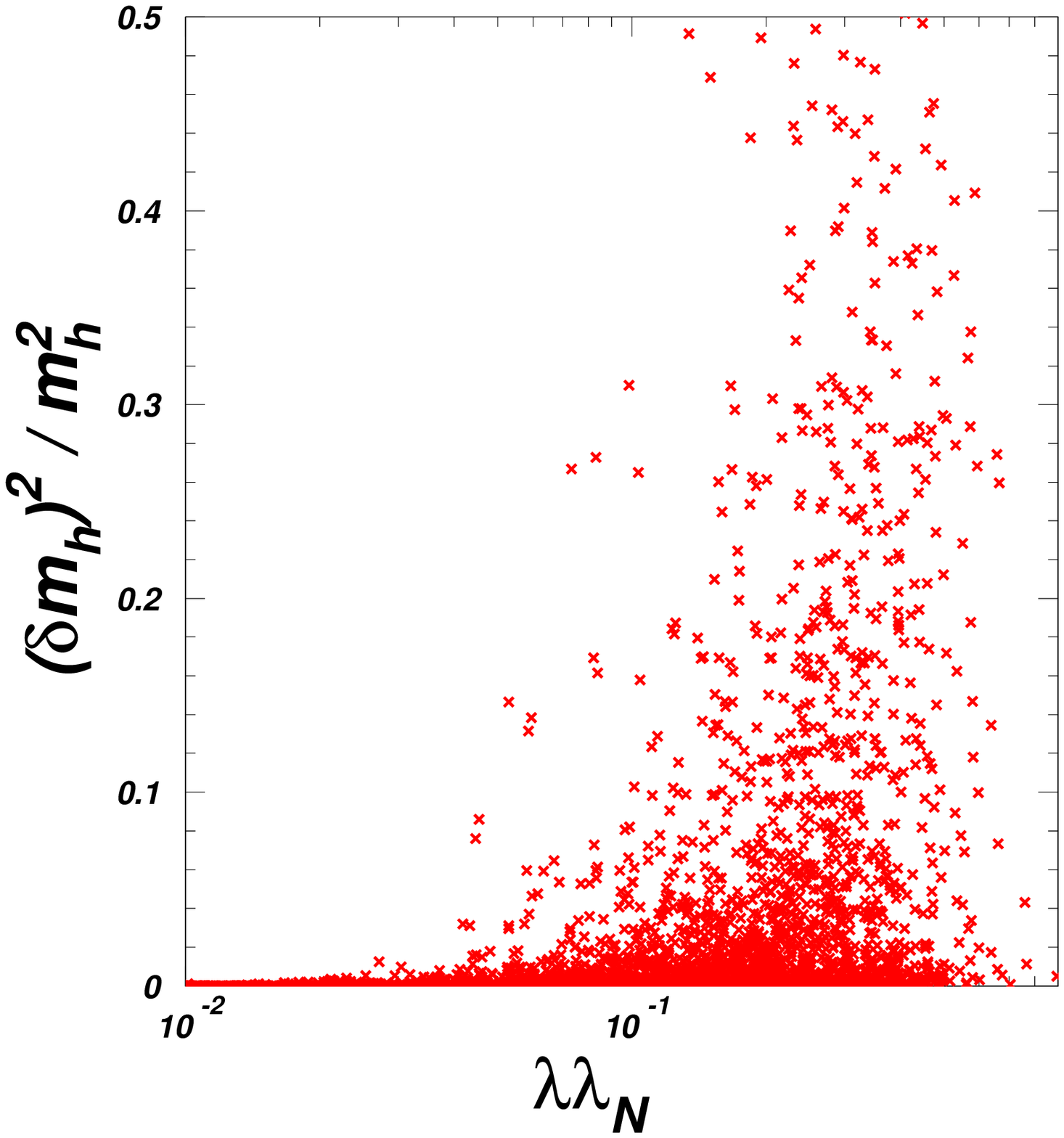}}
\vspace{-.8cm}
\caption{The right-handed neutrino/sneutrino contribution to the
SM-like Higgs boson mass versus $\l \l_N$.}
\label{fig4}
\end{figure}
%%%%%%%%%%%%%%%%%%%%%%%%%%%%%%%%%%%%%%%%%%%
In our numerical calculation we concentrate
on the SM-like Higgs boson which is the lightest CP-even Higgs boson
dominated by the Higgs doublets. From the superpotential in Eq.
(\ref{superpotential}) we can see that the right-handed
neutrino/sneutrino interacts with the doublet only through the
F-term of the singlet Higgs $S$, and thus the parameter $\l$ will
play an important role in the correction to the Higgs boson mass.
Also, from the superpotential we can also
see that the right-handed neutrino/sneutrino couples to the Higgs
sector through the parameter $\l_N$. So, as $\l_N$ approaching zero,
the right-handed neutrino/sneutrino should decouple form the NMSSM
sector.  To check this numerically,
we scan the parameter space in the range:
\bea
&&0<\l,~\k,~\l_N<1, ~~2<\tan\beta<50,\label{range1} \\
&&0<\mu, M_{\tilde N}<1{\rm ~TeV},
~~-1{\rm ~TeV}<A_\l, A_\k, A_N<1{\rm ~TeV},\label{range2}
\eea
Note that in the calculation of the Higgs mass spectrum
we chose to use $\mu$ ($=\lambda v_s$) as an input parameter because
it is commonly used in the NMSSM phenomenology studies
and the relevant numerical packages.
Also we note that
$\lambda$ and $\lambda_N$ may be rather constrained
(e.g., $\lambda$ at weak scale must be below 0.7)
if we require perturbativity of the theory up to the grand
unification scale \cite{zerwas}. Of course, if we just treat
NMSSM as a low energy effective theory, such a stringent
perturbativity constraint will be much relaxed.

The correction to the Higgs boson mass versus
product of $\l$ and $\l_N$ is shown in Fig.\ref{fig4}.
From the figure we can see that when the product of $\l$ and $\l_N$
approaches to
zero, the correction will approach zero; when $\l\l_N$ is at order 1,
the right-handed neutrino/sneutrino will alter the mass
significantly. Thus, if $\l$ and $\l_N$ is not small, then the right-handed
neutrino/sneutrino contribution to the Higgs boson mass must be
taken into account.

\begin{figure}[hbtp]
\scalebox{0.48}{\epsfig{file=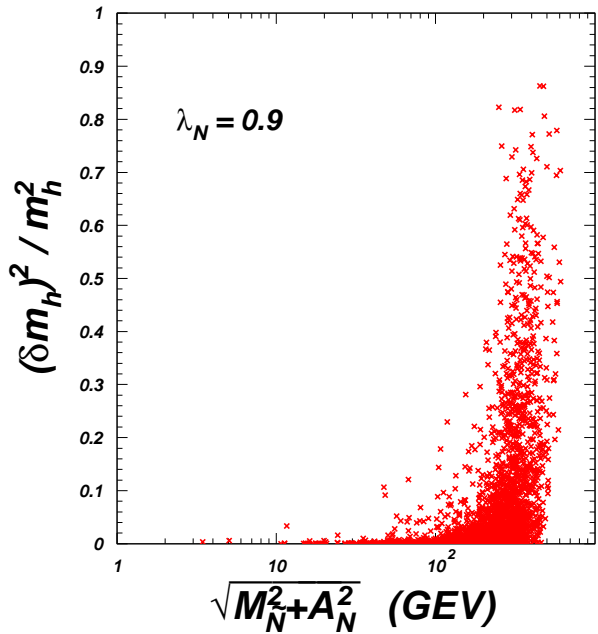}}
\vspace{-.6cm}
\caption{The right-handed neutrino/sneutrino contribution to the
SM-like Higgs boson mass versus $\sqrt{M_{\tilde N}^2+A_N^2}$.}
\label{fig5}
\end{figure}

Now we check the SUSY limit in the right-handed neutrino/sneutrino
sector. From Eq. (\ref{rnspc}) we can see that with $M_{\tilde N}$ and $A_N$
approaching zero, the right-handed neutrino/sneutrino sector has a SUSY
limit for $\k v_s^2 = \l v_u v_d$. In our second scan, we assume the
relation  $\k v_s^2 = \l v_u v_d$ and let the parameter
$\l,~\k,~\tan\beta,~A_\l,~A_\k$, $M_{\tilde N}$ and $A_N$
vary randomly in range as in Eqs. (\ref{range1},~\ref{range2}),
only fixing $\l_N = 0.9$.
The results are shown in Fig. \ref{fig5}.
The results show that with $\sqrt{M_{\tilde N}^2+A_N^2}$ approaching
zero, the Higgs mass correction approaches zero, which confirms the
SUSY limit.
%%%%fig.5 %%%%%%%%%%%%%%%%%%%%%%%%%%%%%%%%%%%%%%%%%%%
%%%%%%%%%%%%%%%%%%%%%%%%%%%%%%%%%%%%%%%%%%%%%%%%%%%

%%%%fig.6 %%%%%%%%%%%%%%%%%%%%%%%%%%%%%%%%%%%%%%%%%%%
\begin{figure}[hbtp]
\scalebox{0.45}{\epsfig{file=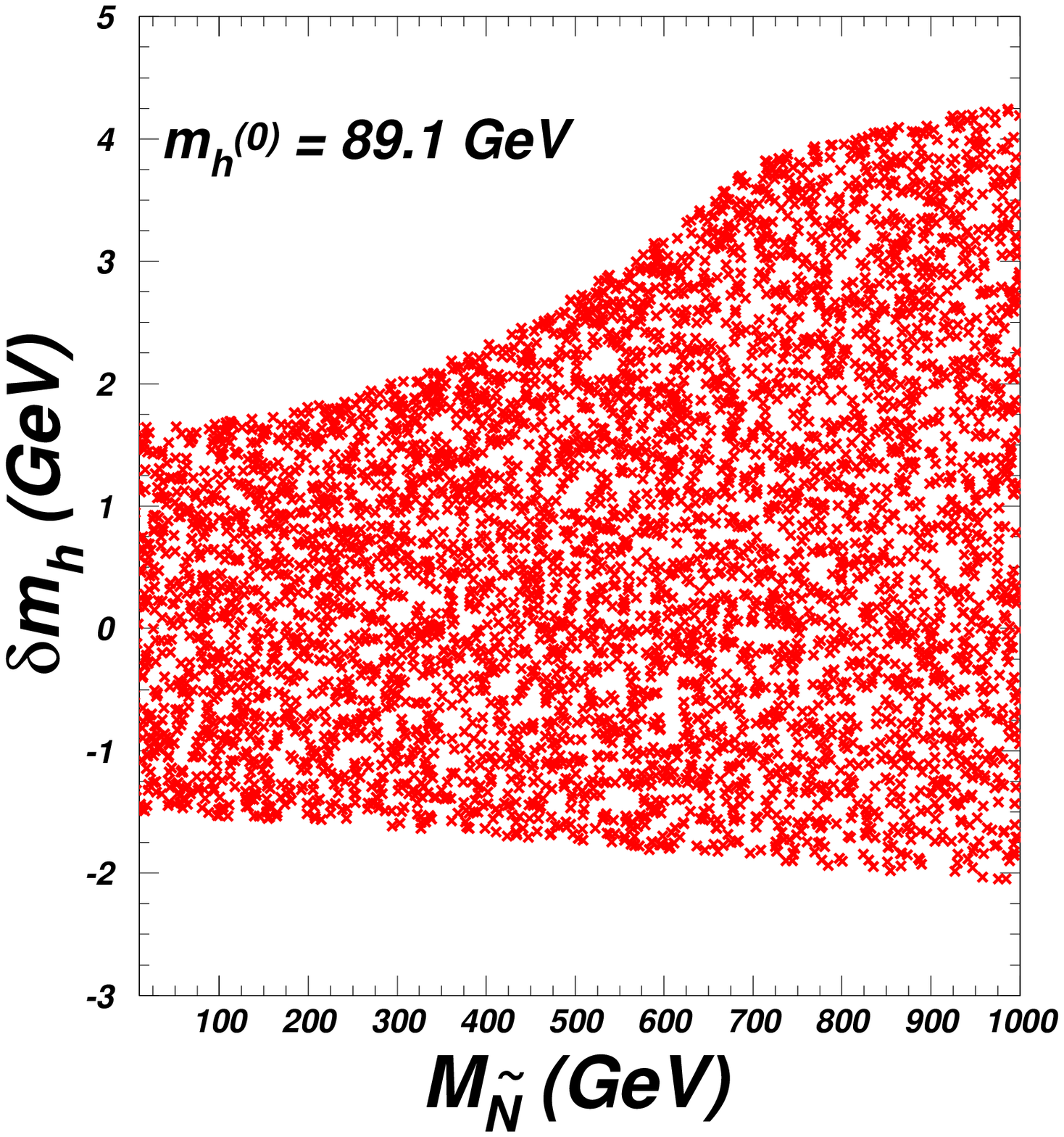}}
\scalebox{0.45}{\epsfig{file=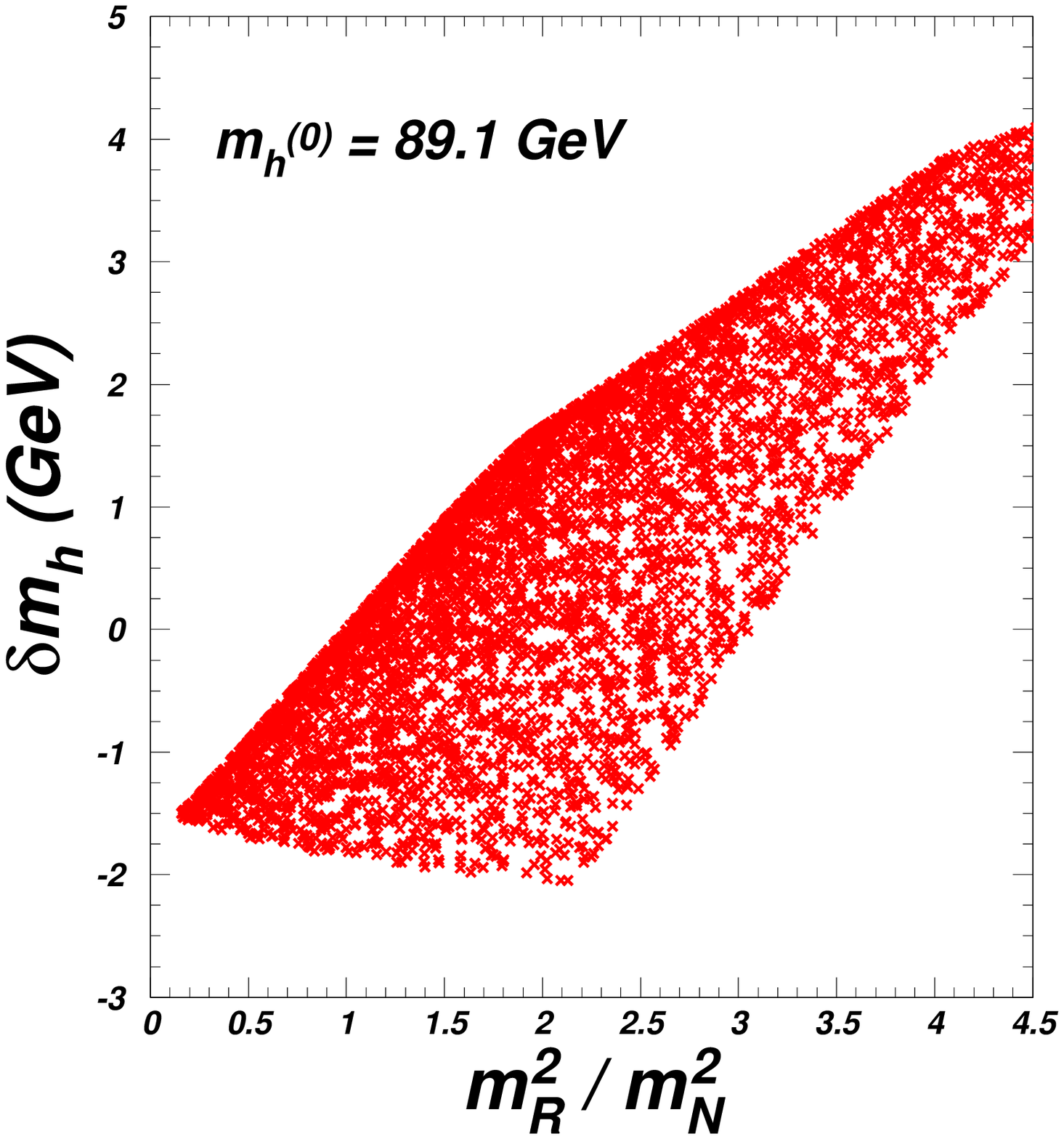}}
\vspace{-.8cm}
\caption{The right-handed neutrino/sneutrino contribution to the
SM-like Higgs boson mass versus the sneutrino soft mass $M_{\tilde N}$
and the ratio $m_R^2/m_N^2$ (for $m_R$ and $m_N$, see Eq. (\ref{rnspc})).}
\label{fig6}
\end{figure}
%%%%%%%%%%%%%%%%%%%%%%%%%%%%%%%%%%%%%%%%%%%
It is well known that the Higgs mass can be enhanced by the
hierarchy between the SM particles and their SUSY partners. In the
right-handed neutrino/sneutrino sector, the mass hierarchy between
sneutrino and neutrino is controlled by the soft parameters
$M_{\tilde N}$ and $A_N$. In order to show the dependence on the
mass splitting, we chose a benchmark point:
\bea
&&\l=0.2,
~\l_N=0.35,
~\k=0.4,
~ \tan\beta=10, \nonumber \\
&& \mu=200 {\rm ~GeV},
~A_\l=300 {\rm ~GeV},
~A_\k=-500  {\rm ~GeV},\label{rang1}
\eea
and scan the other two parameters in the range of
$0<M_{\tilde N}<1{\rm ~TeV}$ and $-1{\rm ~TeV}<A_N<1 {\rm ~TeV}$.
The results are shown in Fig. \ref{fig6}, in which the left panel
shows $\delta m_h$ versus $M_{\tilde N}$ and the right panel shows
$\delta m_h$ versus $m_R^2/m_N^2$. From this figure we can see that
as $M_{\tilde N}$ increases (the mass slitting between sneutrino and
neutrino also increases as shown in Eq. (\ref{rnspc})), the mass
correction increases.

%%%%fig.7 %%%%%%%%%%%%%%%%%%%%%%%%%%%%%%%%%%%%%%%%%%%
\begin{figure}[hbtp]
\scalebox{0.5}{\epsfig{file=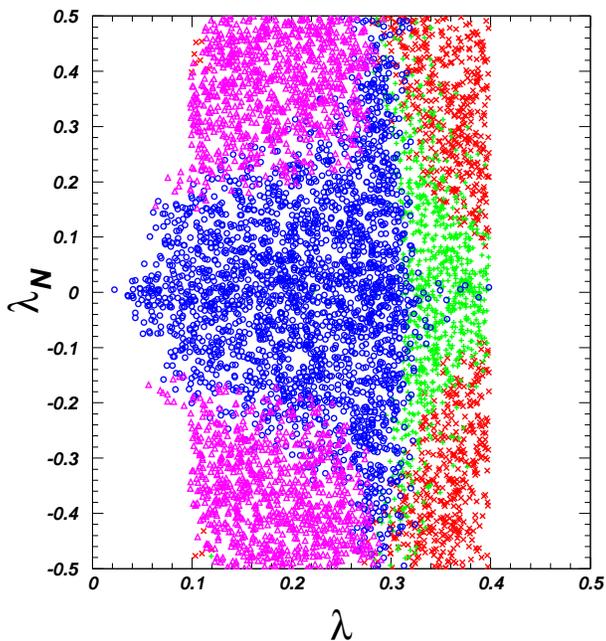}}
\vspace{-.8cm}
\caption{The right-handed neutrino/sneutrino contribution to the
SM-like Higgs boson mass shown in the plane of $\l_N$ versus 
$\l$. Here we scan the parameter set ($\l,~\l_N,~M_{\tilde N},~A_N$) 
while fix other parameters as listed in Eq. (\ref{rang1}). 
The red '$\times$' are for $\delta m_h < -1 {\rm ~GeV}$,
the green '$+$' for $-1 {\rm ~GeV}<\delta m_h < 0$,
 the blue '$\circ$' for $0 < \delta m_h < 1 {\rm ~GeV}$,
 and the magenta '$\triangle$' for $\delta m_h > 1 {\rm ~GeV}$.}
\label{fig7}
\end{figure}
%%%%%%%%%%%%%%%%%%%%%%%%%%%%%%%%%%%%%%%%%%%
From the above results we see that the right-handed neutrino/sneutrino 
can either enhance or reduce the Higgs boson mass. 
Since the parameter space is multi-dimensional (9 input parameters),
we perform an intensive scan to try to figure out what parameter(s) 
determine the sign of the correction. 
We scan the parameter set ($\l,~\l_N,~M_{\tilde N},~A_N$) while 
fix other parameters as listed in Eq. (\ref{rang1}). 
The results are shown in Fig. \ref{fig7}. 
We see that the parameter $\lambda$ plays the most important role 
in this aspect although it cannot solely determine the sign.  
For a large value of $\lambda$ the sign of the correction tends to 
be negative. Clearly, the sign is not sensitive to $\lambda_N$. 
We also checked that the sign is not sensitive to other parameters.

\subsection{Higgs mass with all loop corrections under current experimental constraints}
In the preceding section we only considered the loop corrections
from the right-handed neutrino/sneutrino. Of course, the loop
corrections from other particles (especially the top and stop)
should also be taken into account. In our following numerical study,
we include all available loop corrections by using the package
NMSSMTools \cite{NMSSMTools}. Since the right-handed
neutrino/sneutrino is a gauge singlet, it will not change the Higgs
decay or the annihilation of the dark matter. So, we just add the
right-handed neutrino/sneutrino correction to the Higgs boson mass
in the NMSSMTools. Then we scan the NMSSM parameter space in the
range: \bea
&& 0<\l,~k<1,~~2<\tan\beta<50, \nn\\
&& 0 <(\mu,M_1=M_2/2=M_3/6, ~m_{\tilde Q},
~m_{\tilde t}=m_{\tilde b}=m_{\tilde \tau}=m_{\tilde \mu})<1{\rm ~TeV},\nn \\
&& -1{\rm ~TeV}<(A_\l,~A_\k,~A_t=A_b=A_\tau=A_\mu) <1{\rm ~TeV}.
\eea
For the neutrino/sneutrino sector,
we set $\l_N=0.5$ and scan $M_{\tilde N},~A_N$ in the range
\beq
0< M_{\tilde N}<1{\rm ~TeV},~~-1{\rm ~TeV}< A_N<1{\rm ~TeV}.
\eeq
In our scan we consider the following experimental constraints
\cite{Nakamura:2010zzi}:
(1) We require the lightest neutralino $\tilde{\chi}^0_1$ to account for
the dark matter relic density $0.105 < \Omega h^2 < 0.119$;
(2) We require the SUSY contribution to explain the deviation of the muon
$a_\mu$, i.e., $a_\mu^{\rm exp} - a_\mu^{\rm SM} = ( 25.5 \pm 8.0 ) \times
10^{-10}$ at $2 \sigma$ level; (3) The LEP-I bound on the invisible
$Z$-decay, $\Gamma(Z \to \tilde{\chi}^0_1 \tilde{\chi}^0_1) < 1.76$
MeV, and the LEP-II upper bound on $\sigma(e^+e^- \to
\tilde{\chi}^0_1 \tilde{\chi}^0_i)$, which is $5 \times 10^{-2}~{\rm
pb}$ for $i>1$, as well as the lower mass bounds on the sparticles from
the direct searches at LEP and the Tevatron; (4) The constraints from
the direct search for the Higgs bosons at LEP-II, including the decay
modes $h \to h_1 h_1, a_1 a_1 \to 4 f$, which limit all possible
channels for the production of the Higgs bosons; (5) The constraints
from $B$-physics observables like $B \to X_s \gamma$, $B_s \to
\mu^+\mu^-$, $B^+ \to \tau^+ \nu$, $\Upsilon \to \gamma a_1 $, the
$a_1$--$\eta_b$ mixing and the mass difference $\Delta M_d$ and $\Delta
M_s$; (6) The newest results for Higgs, top and stop results
of the LHC.
These constraints have been encoded in the package NMSSMTools \cite{NMSSMTools}.
In addition to the above experimental limits, we also consider the
constraint from the stability of the Higgs potential, which requires
that the physical vacuum of the Higgs potential with non-vanishing
vevs of Higgs scalars should be lower than any local minima.
%%%%fig.8 %%%%%%%%%%%%%%%%%%%%%%%%%%
\begin{figure}[hbtp]
\scalebox{0.45}{\epsfig{file=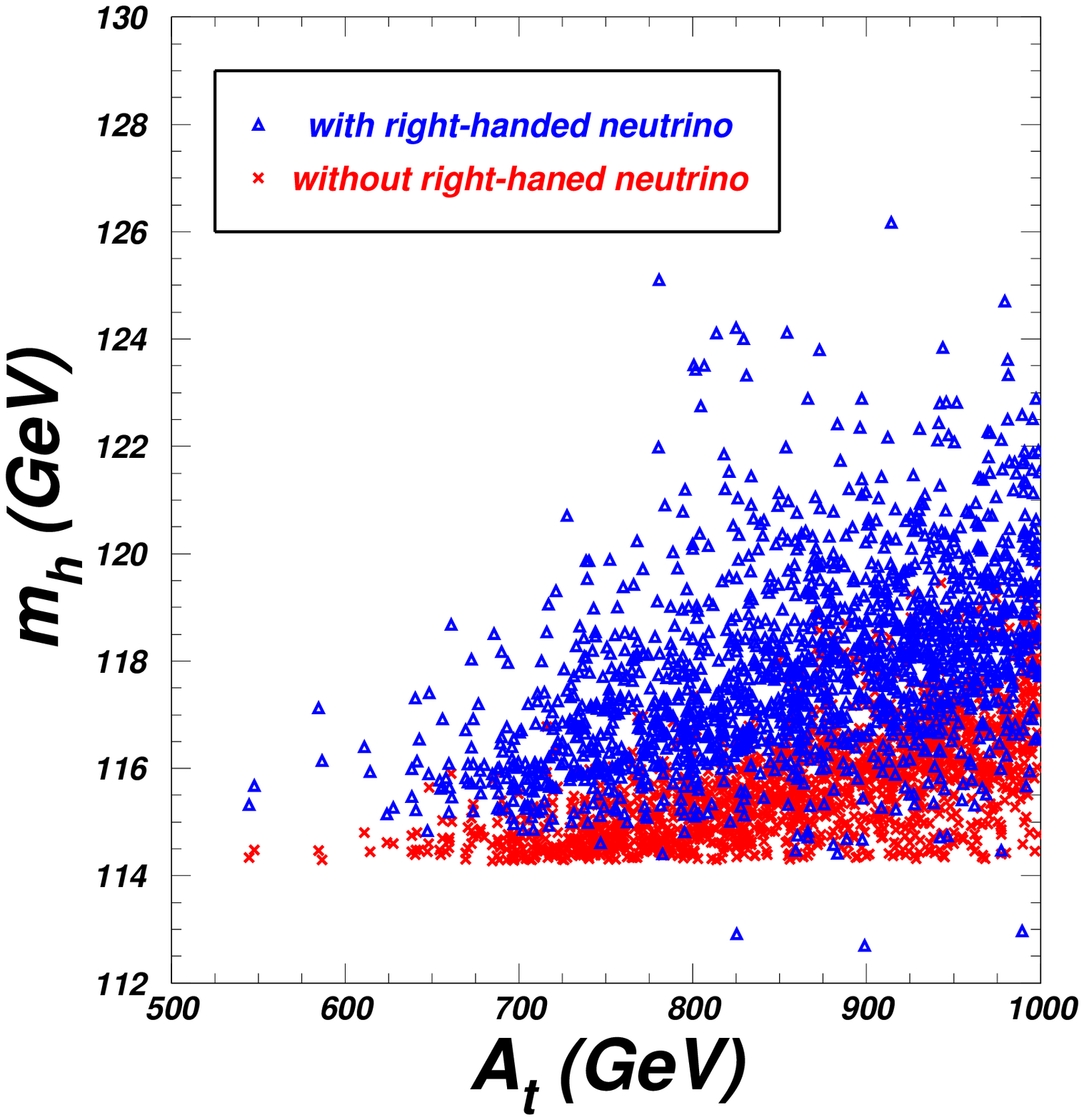}}
\scalebox{0.45}{\epsfig{file=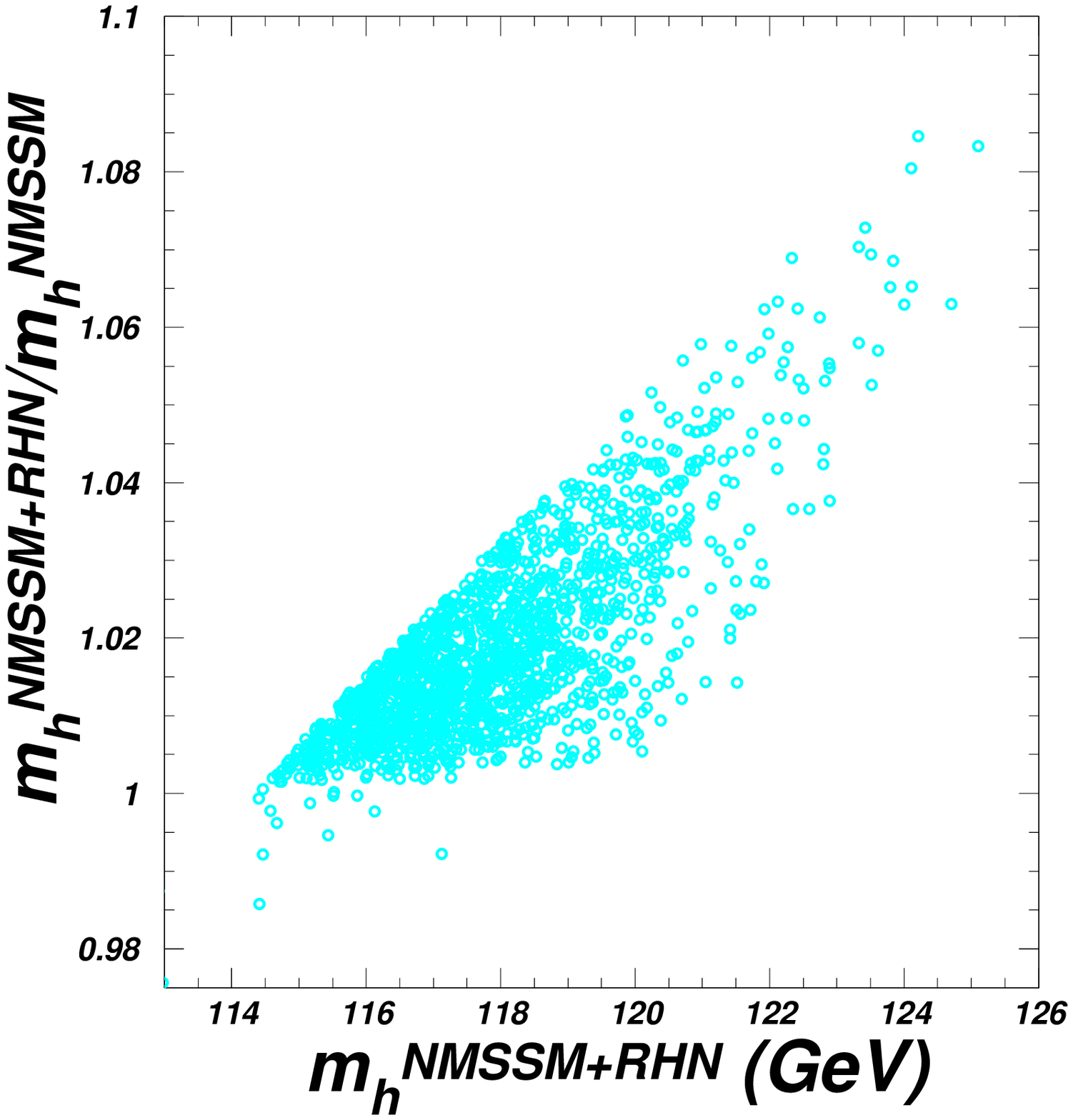}}
\vspace{-.6cm}
\caption{The left panel shows the loop-corrected mass of the SM-like Higgs
with or without the right-handed neutrino/sneutrino
contribution, while the right panel shows the ratio $m_h^{\rm NMSSM+RHN}/m_h^{\rm NMSSM}$
with $m_h^{\rm NMSSM+RHN}$ ($m_h^{\rm NMSSM}$) denoting 
 the SM-like Higgs mass with (without) the right-handed neutrino/sneutrino contribution.}
\label{fig8}
\end{figure}
%%%%%%%%%%%%%%%%%%%%%%%%%%%%%%%%%%%%

The numerical results of our scan are shown in Fig. \ref{fig8} in
which we show the SM-like Higgs mass versus the tri-linear parameter
$A_t$ in the left panel, and the ratio of $m_h^{\rm NMSSM+RHN}/m_h^{\rm NMSSM}$
versus $m_h^{\rm NMSSM+RHN}$ in the right panel. Again we see
that the contribution of the right-handed
neutrino/sneutrino is sizable, which helps to push up the SM-like
Higgs boson mass and thus makes the NMSSM more natural.

Note that from Figs.6 and 7 we see that the correction to the Higgs mass 
can be positive or negative, depending on the parameter space (the most
sensitive parameter is $\lambda$). However, under current experimental constraints
the results in Fig.8 show that in the major part of the survived parameter
space the correction is positive. The reason is that the parameter samples
which give negative corrections are hard to survive the current  
experimental constraints (especially the Higgs mass lower bound given by
LEP-II).   

\section{summary}
\label{sec:sum} In order to have massive neutrinos, the right-handed
neutrino/sneutrino superfield must introduced in SUSY. In the
framework of NMSSM such an extention will dynamically lead to a
TeV-scale Majorana mass for the right-handed neutrino. Further,
through the couplings $SNN$ and $SH_uH_d$, the SM-like Higgs boson
can naturally couple with such TeV-scale right-handed
neutrino/sneutrino. As a result, the right-handed neutrino/sneutrino
may significantly contribute to the Higgs boson mass. In this work
we performed an explicit calculation and found that the Higgs boson
mass can indeed be sizably altered by the right-handed
neutrino/sneutrino. Such new contribution can help to push up the
SM-like Higgs boson mass and thus make the NMSSM more natural.

\section*{ACKNOWLEDEGMENT}
This work was supported in part by NSFC
(No. 11005006, 11275245, 10821504 and 11135003) and
by Ri-Xin Foundation of BJUT  from China.

\section*{Appendix}
Here we list the analytical renormalized formula for the elements
of the Higgs mass matrix. Although they can be found in Ref. \cite{Ender:2011qh},
we checked them and modified them according to our convention.
Note that $\beta_B$ denotes the tree-level $\beta$
and the $c_X,~s_X,~t_X$ denote respectively
$\cos X$, $\sin X$ and $\tan X$.

The scalar $3\times 3$ mass matrix $M_S^2$ in the basis
$h^S=(H_u,H_d,S)^T$ is given by the entries $M^2_{S_{ij}}=
M^2_{S_{ji}}$ ($i,j=1,2,3$) with
\begin{align} \nonumber
M^2_{S_{11}} &=\frac{e c_{\beta} c_{\beta_B}}{2 M_W s_W
  c^2_{\Delta\beta}}  [-t_{H_d} s_{\beta_B} t_{\beta_B} + t_{H_u}
s_{\beta_B} (t_{\beta} t_{\beta_B} + 2)]
\\& \quad\
+ \frac{c_\beta^2 }{c^2_{\Delta\beta}} [M_{H^\pm}^2+(M_Z^2 t_\beta^2 -M_W^2)
  c^2_{\Delta\beta}]
+ \frac{2 \lambda^2 M_W^2 s_W^2 c^2_{\beta}}{e^2},
\displaybreak[3] \\
M^2_{S_{12}} &=
 \frac{e c_\beta c^2_{\beta_B} }{2 M_W  s_W c^2_{\Delta\beta}}
 [t_{H_d} t_{\beta} t^2_{\beta_B} + t_{H_u}]
- \frac{s_{\beta} c_{\beta} }{c^2_{\Delta\beta}} [ M_{H^\pm}^2 +(M_Z^2-M_W^2)
  c^2_{\Delta\beta} ]
+ \frac{\lambda^2 M_W^2 s_W^2 s_{2 \beta}}{e^2}
\displaybreak[3],\\\nonumber
M^2_{S_{13}} &=  \nonumber
\frac{c^2_{\beta} c^2_{\beta_B} }{\sqrt{2}v_s c^2_{\Delta\beta}} [ t_{H_d} t_{\beta} t^2_{\beta_B} +
  t_{H_u} ]
+ \frac{\sqrt{2} M_W s_W s_\beta c^2_\beta }{e v_s c^2_{\Delta\beta}} [M_W^2
c^2_{\Delta\beta} -M_{H^\pm}^2]
\\& \quad\
+\frac{\sqrt{2}\lambda M_W s_W c_\beta v_s }{e} [2 \lambda t_\beta - \kappa ]
+ \frac{-2\sqrt{2} \lambda^2 M_W^3 s_W^3 s_\beta c^2_{\beta}}{e^3 v_s},
\displaybreak[3]\\
M^2_{S_{22}} &= \frac{e c_\beta c^2_{\beta_B} }{2 M_W s_W
  c^2_{\Delta\beta}} [t_{H_d} (2 t_\beta t_{\beta_B} +1) -t_{H_u} t_\beta]
\\& \quad\
+ \frac{s^2_{\beta} }{c^2_{\Delta\beta}} [M_{H^\pm}^2+(M_Z^2 t^{-2}_\beta-M_W^2)
  c^2_{\Delta\beta}]
+ \frac{2 \lambda^2 M_W^2 s_W^2s^2_\beta}{e^2},
\displaybreak[3]\\
M^2_{S_{23}} &= \frac{s_\beta c_\beta c^2_{\beta_B} }{\sqrt{2} v_s
  c^2_{\Delta\beta}} [t_{H_d} t_{\beta} t^2_{\beta_B} + t_{H_u}]
+ \frac{\sqrt{2} M_W s_W s_\beta^2 c_\beta }{e v_s c^2_{\Delta\beta}} [M_W^2
c^2_{\Delta\beta} -M_{H^\pm}^2]
\\& \quad\
+ \frac{\sqrt{2}\lambda M_W s_W c_\beta v_s }{e} [2 \lambda - \kappa t_\beta]
+\frac{-2\sqrt{2} \lambda^2 M_W^3 s_W^3 s^2_{\beta} c_{\beta}}{e^3 v_s}
\displaybreak[3], \\ \nonumber
M^2_{S_{33}} &=  \nonumber
\kappa A_\kappa v_s +4 \kappa^2 v_s^2+\frac{t_{H_s}}{\sqrt{2}v_s}
+ \frac{M_W s_W s_\beta c_\beta^2 }{e^2 v_s^2 c^2_{\Delta\beta}}
[2 M_{H^\pm}^2 M_W s_W s_\beta - e (t_{H_d}  t_\beta s^2_{\beta_B} +
t_{H_u}  c^2_{\beta_B})]
\nn \\ & \quad\
+ \frac{M_W^2 s_W^2 s_{2\beta}}{2 e^4 v_s^2} [2\lambda^2 M_W^2 s_W^2
s_{2\beta} - 2\kappa \lambda e^2 v_s^2 - M_W^2 e^2 s_{2\beta}] \quad .
\end{align}

The entries $M^2_{P_{ij}}= M^2_{P_{ji}}$ ($i,j=1,2,3$) of the
pseudoscalar $3\times 3$ mass matrix $M_P^2$  in the basis
$h^P=(a,a_s,G)^T$ read
\begin{align}
M^2_{P_{11}} &=\frac{2 \lambda^2 M_W^2 s_W^2 c^2_{\Delta \beta}}{e^2}
+M_{H^\pm}^2 -M_W^2 c^2_{\Delta \beta},
\displaybreak[3]\\\nonumber
M^2_{P_{12}} &=
\frac{M_W  s_W s_{2 \beta}}{\sqrt{2} e v_s c_{\Delta \beta}} [M_{H^\pm}^2 -
M_W^2 c_{\Delta\beta}^2]
-\frac{c_{\beta} c^2_{\beta_B}}{\sqrt{2} v_s c_{\Delta \beta}} [t_{H_u} +
t_{H_d} t_{\beta} t^2_{\beta_B}]
\\& \quad\
+ \frac{\lambda M_W s_W c_{\Delta \beta}}{\sqrt{2} e^3 v_s} [2\lambda M_W^2
s_W^2 s_{2\beta} - 6\kappa e^2 v_s^2], \\
M^2_{P_{13}} &= M_{H^\pm}^2 t_{\Delta \beta}
+\frac{M_W^2 s_{2 \Delta \beta}}{2e^2} [2 \lambda^2 s_W^2 - e^2]
+\frac{e c_{\beta_B}}{2 M_W s_W c_{\Delta \beta}} [t_{H_d} t_{\beta_B}
-t_{H_u}], \displaybreak[3]\\
\nonumber
M^2_{P_{22}} &=-3 A_\kappa \kappa v_s + \frac{t_{H_s}}{\sqrt{2} v_s}
- \frac{M_W s_W s_\beta c_\beta^2 c^2_{\beta_B}}{e^2 v_s^2 c_{\Delta \beta}^2}
 [t_{H_u}+ t_{H_d} t_\beta t^2_{\beta_B}]
\\& \quad\
+ \frac{M_W^2 s_W^2 s_{2\beta}^2}{2 e v_s^2 c_{\Delta\beta}^2}
[M_{H^\pm}^2-M_W^2 c_{\Delta \beta}^2]
+ \frac{\lambda M^2_W s^2_W s_{2\beta}}{e^4 v_s^2} [\lambda M_W^2
s_W^2 s_{2\beta} + 3\kappa e^2 v_s^2],
\displaybreak[3]\\ \nonumber
M^2_{P_{23}} &= \frac{M_W s_W s_{2\beta}}{2\sqrt{2}ev_s c_{\Delta \beta}} [2
M_{H^\pm}^2 t_{\Delta \beta} - M_W^2 s_{2\Delta \beta}]
-\frac{c_{\beta} c^2_{\beta_B} t_{\Delta \beta} }{\sqrt{2}v_s c_{\Delta
    \beta}} [t_{H_u} + t_{H_d} t_{\beta} t^2_{\beta_B}]
\\& \quad\
+ \frac{\lambda M_W s_W s_{\Delta \beta}}{\sqrt{2} e^3 v_s} [2\lambda M_W^2
s_W^2 s_{2\beta} - 6\kappa e^2 v_s^2],
\displaybreak[3]\\ \nonumber
M^2_{P_{33}} &= M_{H^\pm}^2 \tan ^2{\Delta \beta}  + \frac{M_W^2 \sin
  ^2{\Delta \beta}}{e^2} [2 \lambda^2 s_W^2 - e^2]
\\& \quad\
+\frac{e }{2 M_W s_W c^2_{\Delta \beta}} [t_{H_d} c_{\beta - 2
  \beta_B} -t_{H_u} s_{\beta - 2 \beta_B}] \quad .
\end{align}

\end{document}